\documentclass[prd,twocolumn,floatfix,superscriptaddress,nofootinbib]{revtex4-1} %showpacs,
\usepackage{graphicx}
\usepackage{amssymb}
\usepackage{dcolumn}
\usepackage{amsmath}
\usepackage{bm}
\usepackage{textcomp}
\usepackage{epstopdf}
\usepackage{color}
\usepackage{ulem}
\usepackage[toc,page,title,titletoc,header]{appendix}
\usepackage[colorlinks,
            linkcolor=blue,
            anchorcolor=blue,
            citecolor=blue
            ]{hyperref}
\usepackage{txfonts}
\usepackage{float}
\usepackage{multirow}
\begin{document}
\title{Deep learning stochastic processes with QCD phase transition}

\begin{abstract}
It is non-trivial to recognize phase transitions and track dynamics inside a stochastic process because of its intrinsic stochasticity. In this paper, we employ the deep learning method to classify the phase orders and predict the damping coefficient of fluctuating systems under Langevin's description. As a concrete set-up, we demonstrate this paradigm for the scalar condensation in QCD matter near the critical point, in which the order parameter of chiral phase transition can be characterized in a $1+1$-dimensional Langevin equation for $\sigma$ field.
In a supervised learning manner, the Convolutional Neural Networks(CNNs) accurately classify the first-order phase transition and crossover based on $\sigma$ field configurations with fluctuations. Noise in the stochastic process does not significantly hinder the performance of the well-trained neural network for phase order recognition. For mixed dynamics with diverse dynamical parameters, we further devise and train the machine to predict the damping coefficients $\eta$ in a broad range. The results show that it is robust to extract the dynamics from the bumpy field configurations.

\end{abstract}

\author{Lijia Jiang}
%\email{jiang@fias.uni-frankfurt.de}
\affiliation{Frankfurt Institute for Advanced Studies, Ruth Moufang Strasse 1, D-60438,
Frankfurt am Main, Germany}
\author{Lingxiao Wang}
\affiliation{Frankfurt Institute for Advanced Studies, Ruth Moufang Strasse 1, D-60438,
Frankfurt am Main, Germany}

\author{Kai Zhou}
\email{zhou@fias.uni-frankfurt.de}
\affiliation{Frankfurt Institute for Advanced Studies, Ruth Moufang Strasse 1, D-60438,
Frankfurt am Main, Germany}
\maketitle

\section{Introduction}
% in polishing
The phenomena of phase transition are extensively observed in various many-body systems. Through measuring the thermodynamic quantities such as the susceptibility or heat capacity, the information about the phase transition (e.g., the order of the phase transition, the critical exponent, etc.) could be extracted no matter in classical or quantum systems. In addition to its own intricacy for quantum systems, the situation becomes more complicated in a non-equilibrium dynamical evolution, especially for a stochastic process. The intrinsic randomness breaks the deterministic description for such stochastic dynamics, which hinders our further understanding to those exotic non-equilibrium systems, e.g., cold atoms in a moving optical lattice~\cite{chong:2018observation}, or heavy-quark diffusion in the Quark-Gluon Plasma(QGP)~\cite{vanhees:2008nonperturbative,zhao:2020heavy}.

Deep learning with a hierarchical structure of artificial neural networks is emerging as a novel tool to deal with high-level representations of intricate data~\cite{lecun:2015deep}. With the advancement of hardware and computational power, there are significant progresses of applications of deep learning in an increasing number of fields, such as audio recognition, medical image analysis, computer vision, and board game programs, etc. In these cases, the machine has produced results comparable or even superior to human experts. Recently, the deep learning method is also utilized in the field of physics research~\cite{carleo:2019machine}, such as in nuclear physics~\cite{pang:2019interpretable,huang:2019applications,liu:2019principal,omanakuttan:2020fast,pang:2018equationofstatemeter,du:2020identifying,zhou:2019regressive}, particle physics~\cite{Baldi:2014kfa,Baldi:2014pta,barnard:2017parton,Moult:2016cvt,Radovic:2018dip}, and condensed matter physics~\cite{wang:2016discovering,vannieuwenburg:2017learning,broecker:2017machine,carrasquilla:2017machine,Torlai_2016,Han_2018,wang:2020continuousmixture}. The advantage of deep learning method is that it could help us extracting hidden correlations from complex non-linear physical systems, which might be difficult to tackle in the conventional computation.

Model-free prediction with machine learning on state evolution has been discussed in recent years for nonlinear/chaotic dynamical systems%Start from the point, predicting the stochastic process or even the evolution of chaotic systems becomes possible
~\cite{pathak:2017using,pathak:2018modelfree,fan:2020longterm}.
In the present paper, we explore application of deep learning to detect phase transition and dynamical information in stochastic processes, which would be of great potential application in a variety of fields.
In such problems, unlike the case with thermal equilibrium,
the raw observations (configurations) for the system are nothing but the stochastic time series data. To uncover the phase transition or the dynamical information from limited raw data with stochasticity is in principle challenging but crucial for studying the properties of the dynamical system.
A related paradigm that has been developed for deterministic (with minor stochasticity in some stages) dynamical system~\cite{pang:2018equationofstatemeter,du:2020identifying} is to train a deep neural network supervisedly to identify the phase order of QCD Equation of State (EoS) in heavy-ion collisions.  We generalize the idea further to recognize the phase order and extract the dynamical parameters in a stochastic dynamical process with phase transition. With regard to the effective inputs to the Deep Neural Networks(DNNs), while the final-state particle spectra could be a proper choice for deterministic systems~\cite{pang:2018equationofstatemeter,du:2020identifying}, we feed the event-by-event temporal-spatial scalar field configurations in final stage to the neural network to identify the dynamics, including the phase order and dynamical parameter. Specifically, we design a deep Convolutional Neural Network (CNN) to track the dynamical process with phase transition. It is worth to note that the magnitude of the noise in the dynamical evolution does not hinder the classification ability of the well-trained CNNs noticeably. The neural network have a consistent performance on predicting the phase order of the configuration with unknown fluctuations. Moreover, we utilize the machine to predict the damping coefficient from the configurations in final stage of the stochastic process. In this part, we evolve the system within the crossover scenario with diverse damping coefficients. As the dominate dynamical parameter, the damping coefficient is set inside a limited range for the training data sets. The test on a broader range of the damping achieves a good performance, which suggests that the deep CNNs could help extracting crucial dynamical information in such stochastic Langevin dynamics given just the limited raw states.

%A recurrent neural network (RNN) is a class of artificial neural network where connections between nodes form a directed graph along a temporal sequence. This allows it to exhibit temporal dynamic behavior. The advantage of RNN is the network can use their internal state (memory) to process sequences of inputs. This makes them applicable to tasks such as unsegmented, connected handwriting recognition or speech recognition.

%The Langevin equation is a stochastic differential equation which is generally applied to different kinds of system, in which the degree of freedom typically are the collective variables changing very slowly, compared to the other microscopic variables in the system. The microscopic variables lead to the stochastic nature of the Langevin equation. There are a lot of physical systems can be described by the Langevin dynamics, such as the harmonic oscillator in a fluid, thermal noise in an electrical resistor, and the critical dynamics in different system. Of our interest is the system with critical dynamics.
%One question of both theoretical and experimental interests is if we can identify the order of the phase transition and the related critical exponent of a dynamical system from the information of time-space configurations.

The paper is organized as following: In Sec.~\ref{sec:dl}, we introduce the paradigm that applying deep learning to Langevin dynamics of QCD matter, in which the parameter set-ups and details of the event-by-event simulation in 1+1d Langevin equation are described. In Sec.~\ref{sec:pt}, the Convolutional Neural Networks(CNNs) are adopted to classify the first-order transition and crossover from $\sigma$ field configurations with fluctuations. The stability of the performance is evaluated for data set with various noise. In Sec.~\ref{sec:dynamics}, we prepare mixed configurations with damping coefficient in the range of $(1.0-2.5)~fm^{-1}$ and $(4.6-5.5)~fm^{-1}$ for training the machine to recognize the damping coefficient, however, the predictions to the dynamics are made for a broader range beyond training. It is found that the dynamics recognition from such stochastic process with neural network is robust. In Sec.~\ref{sec:discus}, we summary the main findings in this work and conclude the potentials of our paradigm.

\section{Deep Learning Langevin dynamics}
\label{sec:dl}

In this section, we introduce a deep learning approach to track the stochastic process driven by a Langevin equation, in which the scalar $\sigma$ field evolves as the order parameter of QCD phase transition~\cite{paech:2003hydrodynamics,herold:2013chiral,jiang:2017dynamical,jiang:2018enhancements} near the critical point~\cite{schaefer:2008renormalization,qin:2011phase}. Although the dynamical process is discussed in the context of high energy heavy-ion collision systems, it can be naturally extended into a general non-equilibrium stochastic systems~\cite{pavliotis:2014stochastic} in a broader areas.

\begin{figure}[htbp!]
\center
\includegraphics[width=2.9 in] {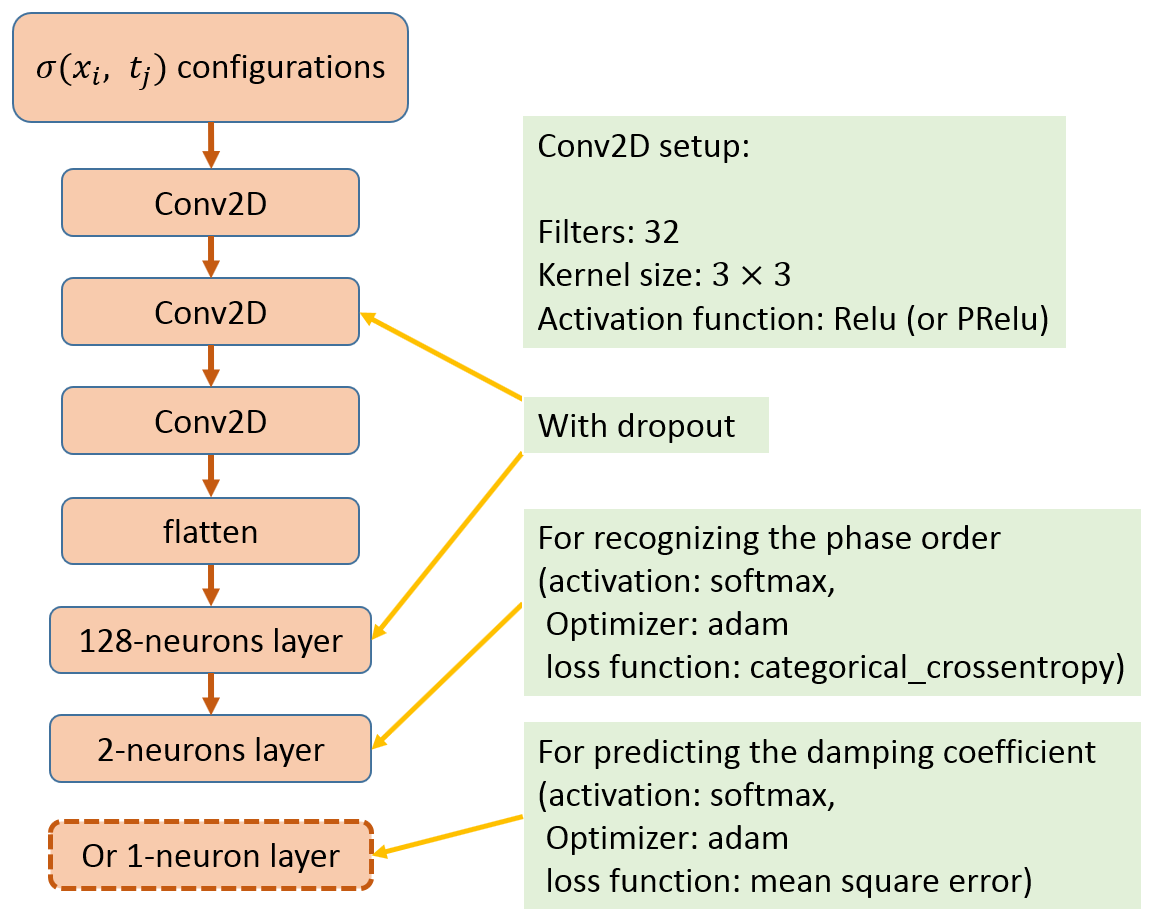}
\caption{The architecture of the deep CNNs for recognizing the phase order and predicting the damping coefficient from the $\sigma$ field configurations. Details of the network structure can be found in the text especially for the damping regression task.}
\label{fig:flowchart}
\end{figure}

In particular, we implement the deep CNNs to explore the dynamical process as the flowchart in Fig.~\ref{fig:flowchart} shown.
To prepare the training data for the CNNs, we numerically solve the event-by-event Langevin equation for the scalar $\sigma$ field. The temporal-spatial information of the field configurations $\sigma(x,t)$ are recorded when the system temperature drops far below the critical temperature $T_c$. To perform supervised learning, the configurations labeled with crossover and first-order phase transition are prepared to be our training set, with which, we train the CNNs to classify the order of the phase transition given configurations only in later stage of the dynamical evolution. As a demonstration of robustness, we test the trained CNNs in various cases with different magnitude of noise. Further we also show that, as the key dynamical information, the damping coefficient $\eta$ can be well recognized via deep CNNs.

%%%%%%%%%%%%%%%%%%%%%%%%%%%%%%%%%%%%%%%%%%%%%%%%%%%%%%%%%%%%%%%%%%%%%%%%%%%%%%
\begin{figure}[htbp!]
\centering
\includegraphics[width=0.99\linewidth]{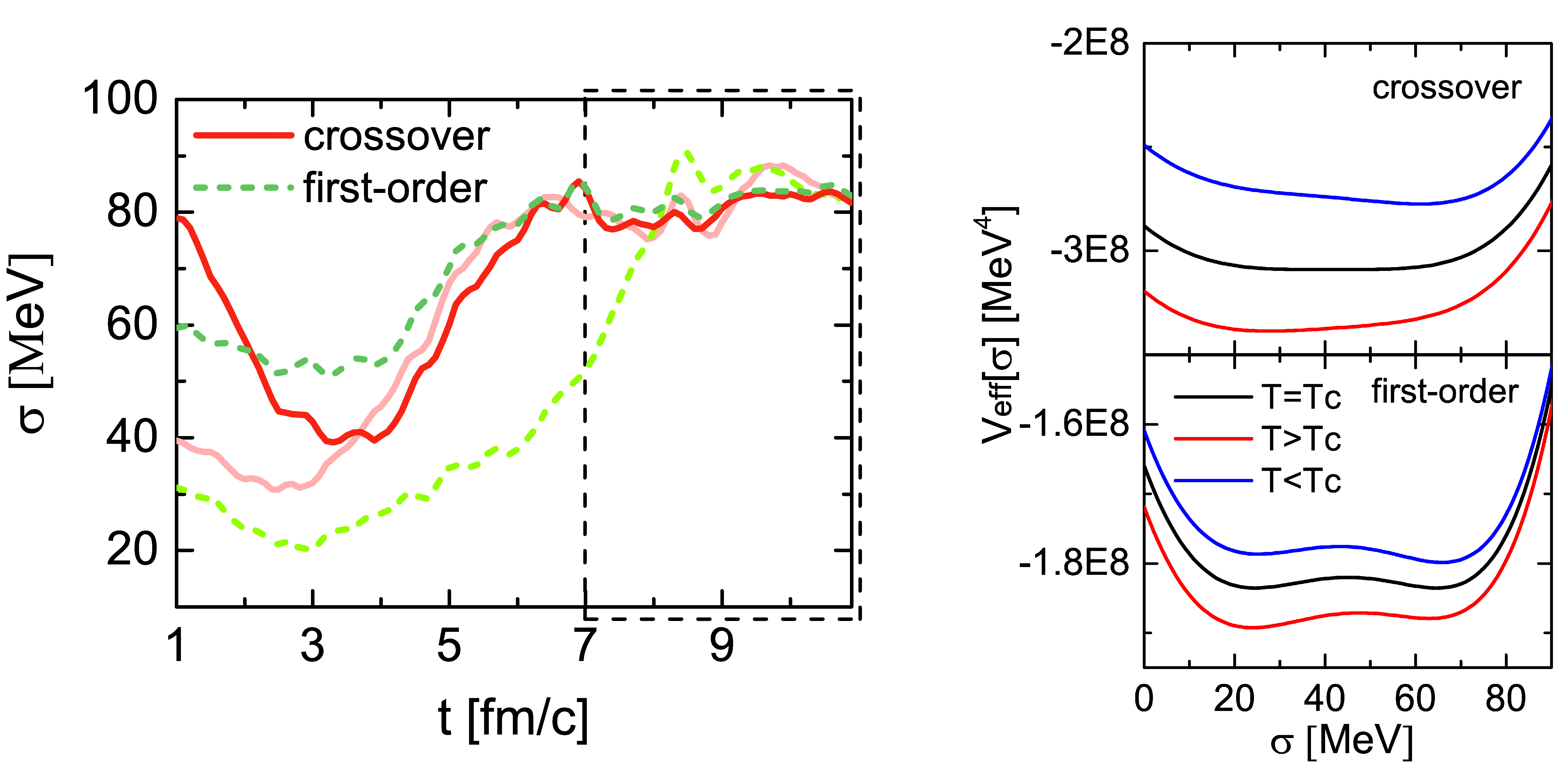}
\caption{Dynamical evolution of scalar $\sigma$ field in different phase transition scenarios with solid lines standing for crossover and dashed lines for first-order phase transition, with the small box on the upper right corner masking the input range we chose for the CNNs. The right figures are the corresponding effective potential which induces the different phase transitions, in which the upper one shows the crossover with temperature across the critical temperature $T_c$, and the bottom one shows the first-order transition.}
\label{fig:potential}
\end{figure}

%%%%%%%%%%%%%%%%%%%%%%%%%%%%%%%%%%%%%%%%%%%%%%%%%%%%%%%%%%%%%%%%%%%%%%%%%%%%%%
With increasing temperature $T$ and/or baryon chemical potential $\mu$, the QCD matter will undergo a phase transition from hadronic matter to the quark gluon plasma (QGP) phase. Further, the QCD transition is conjectured to be a crossover at small chemical potential $\mu$ (and moderately high temperature), and first order at moderate values of $\mu$ (and lower temperature), with a critical point separating the two.
%In the parameter space of temperature $T$ and baryon chemical potential $\mu$， the QCD matter will transfer from the hadron phase into the quark phase with temperature or density increasing.
Although the QCD phase transition is complicated~\cite{fukushima:2011phase,fukushima:2013phase}, its general thermodynamics and phase behaviors could be describe effectively by models such as Nambu--Jona-Lasinio model~\cite{hatsuda:1994qcd,jiang:2013revisiting,wang:2018nambu} Quark-Meson Model~\cite{jungnickel:1996effective,schaefer:2005phase}, or Linear Sigma Model (LSM)~\cite{petropoulos:1999linear}. As a practical example, the effective potential from the LSM model presents a scenario in which the crossover locates at the small chemical potential region and first-order phase transition occurs at the large chemical potential region~\cite{roh:1998chiral}. In Heavy Ion Collisions (HICs) experiments, the hot and dense fireball created sets an extreme dynamical environment where the QCD phase transition can happen~\cite{aoki:2006order,hotqcdcollaboration:2012chirala,bzdak:2020mapping}. To model the phase transition processes in HICs, the Langevin equation is adopted to describe the semi-classical evolution for the long wavelength mode of the $\sigma$ field (for more context and details can see~\cite{nahrgang:2011nonequilibrium,nahrgang:2013impact}),
\begin{equation}
\partial ^{\mu }\partial _{\mu }\sigma \left( t,x\right) +\eta \partial
_{t}\sigma \left( t,x\right) +\frac{\delta V_{eff}\left( \sigma \right) }{%
\delta \sigma }=\xi \left( t,x\right),
\label{eq:langevin}
\end{equation}
where $\eta $ is the damping coefficient, $\xi \left( t,x\right) $ is the noise term, and the effective potential $V_{eff}$ decides the type of phase transition in the stochastic process as shown in Fig.~\ref{fig:potential} (see more details in Ref.~\cite{paech:2003hydrodynamics}).
The terms with $\eta$ and $\xi$ are both from the interaction
between the $\sigma$ field and the thermal background, and satisfy the fluctuation-dissipation theorem: $%
\left\langle \xi \left( t\right) \xi \left( t^{\prime }\right) \right\rangle
\sim \eta \delta \left( t-t^{\prime }\right) $~\cite{nahrgang:2011nonequilibrium}.
In our calculation, $\eta$ is taken as a free parameter while the noise is set as the white noise.
%In the stochastic process, the effective potential $V_{eff}\left( \sigma \right)$ decides the type of phase transition shown in Fig.~\ref{fig:potential} (see more details in Ref.~\cite{paech:2003hydrodynamics}).

%%%%%%%%%%%%%%%%%%%%%%%%%%%%%%%%%%%%%%%%%%%%%%%%%%%%%%%%%%%%%%%%%%%%%%%%%%%%%%
%%%%%%%%%%%%%%%%%%%%%%%%%%%%%%%%%%%%%%%%%%%%%%%%%%%%%%%%%%%%%%%%%%%%%%%%%%%%%%

In a practical implementation, we adopt the method proposed in Ref.~\cite{jiang:2017dynamical} to construct the initial profiles of the $\sigma$ field, according to the probability distribution function $ P\left( \sigma \right) \sim \exp \left( -E \left( \sigma\right) /T\right) $ with the energy function $ E \left[ \sigma\left(\mathbf{x}\right) \right] =\int d^{3}x\left[ \frac{1}{2}\left( \nabla \sigma \left( x\right) \right)^{2}+V_{eff}\left( \sigma \left( x\right) \right) \right] $).
To mimic the realistic dynamical process in HICs where the created fireball expands and cools rapidly, in principle we need to embed the local temperature $T (t,x,y,z)$ and baryon chemical potential $\mu(t,x,y,z)$ into the effective potential.  For simplicity but without loss of generality, we assume the heat bath evolves along trajectories with constant baryon chemical potential, and
the temperature drops down in a Hubble-like way,
\begin{equation}
\frac{T\left( t\right) }{T_{0}}=\left( \frac{t}{t_{0}}\right) ^{-0.45},
\label{eq:temp}
\end{equation}
where $T_{0}~(>T_c)$ is the initial temperature, and $t_{0}=1$ fm is the initial time for the evolution.
With regard to the dynamical evolution of the $\sigma $ field, we set the damping
coefficient $\eta$ to be constant across the evolution with values ranging from $1.0$ to $5.5$ $fm^{-1}$.
%which is consistent in the following calculations.
As for the details of the numerical set-up, we simulate the evolution of $\sigma$ field in $1$-dimensional space with range $L=6.0$ fm, and the spatial grid size $dx = 0.2$ fm. The duration of the evolution is $16$ fm at most with the temporal step size $dt=0.1$ fm/c. With the above set-ups and initial profiles, the $\sigma$ field was evolved according to Eq.~\eqref{eq:langevin} on an event-by-event basis.
As a practical choice, the configurations from later episode well after phase transition with $4$ fm duration of the $\sigma$ field are censored as the input data set. Thus, the input is in $t\in[7, 11]\,\text{fm}^{-1}$, or in the last 40 time-steps from the evolution, where the ambient temperature is already much lower than $T_c$, ensured that the potential phase transition already happened. Therefore, the prepared input configuration contains $N= 40\times 30 = 1200$ pixels in each event.

\section{Recognizing phase transition in stochastic process}
\label{sec:pt}
In this section we first demonstrate that the QCD phase order could be recognized from the stochastic process by a deep CNN. Despite that the evolution is simulated in 1+1-dimensional space for our Langevin systems, its stochastic nature induces elusiveness because of the randomness from interactions. For time series data analysis, long short-term memory (LSTM) neural networks are routinely adopted. However, it is laborious to capture the dynamics in such simplified stochastic processes~\cite{yeo:2019deep} with LSTM networks. As a succinct alternative, we adopt the CNNs to classify the QCD phase order from noisy configurations, since the deep CNN could unearth sufficient correlations in high dimensional data~\cite{dassarma:2019machine}. The main architecture of the deep CNNs is shown in Fig.~\ref{fig:flowchart}. The configurations of $\sigma$ field with a  $40\times 30$ temporal-spatial ``resolution" are fed into the deep CNN as images, followed by 3 convolutional layers with ReLU activation functions as the core structure. For each layer, there are 32 filters with size $3\times3$. To avoid over-fitting, Dropout is applied after the second convolutional layer. After all CNN layers, the outputs are flattened and further fed into a 128-neurons fully connected layer, with ReLU activation function and Dropout follows.
%The final layer is full-connected with 2-neuron, and its outputs derive 2 classes after passing a Softmax activation function.
To tackle this binary classification task, the final output layer of the model is another fully connected layer with softmax activation and 2 neurons to indicate the two phase transition classes.

%%%%%%%%%%%%%%%%%%%%%%%%%%%%%%%%%%%%%%%%%%%%%%%%%%%%%%%%%%%%%%%%%%%%%%%%%%%%%%
\begin{figure}[htbp!]
\centering
\includegraphics[width=0.99\linewidth]{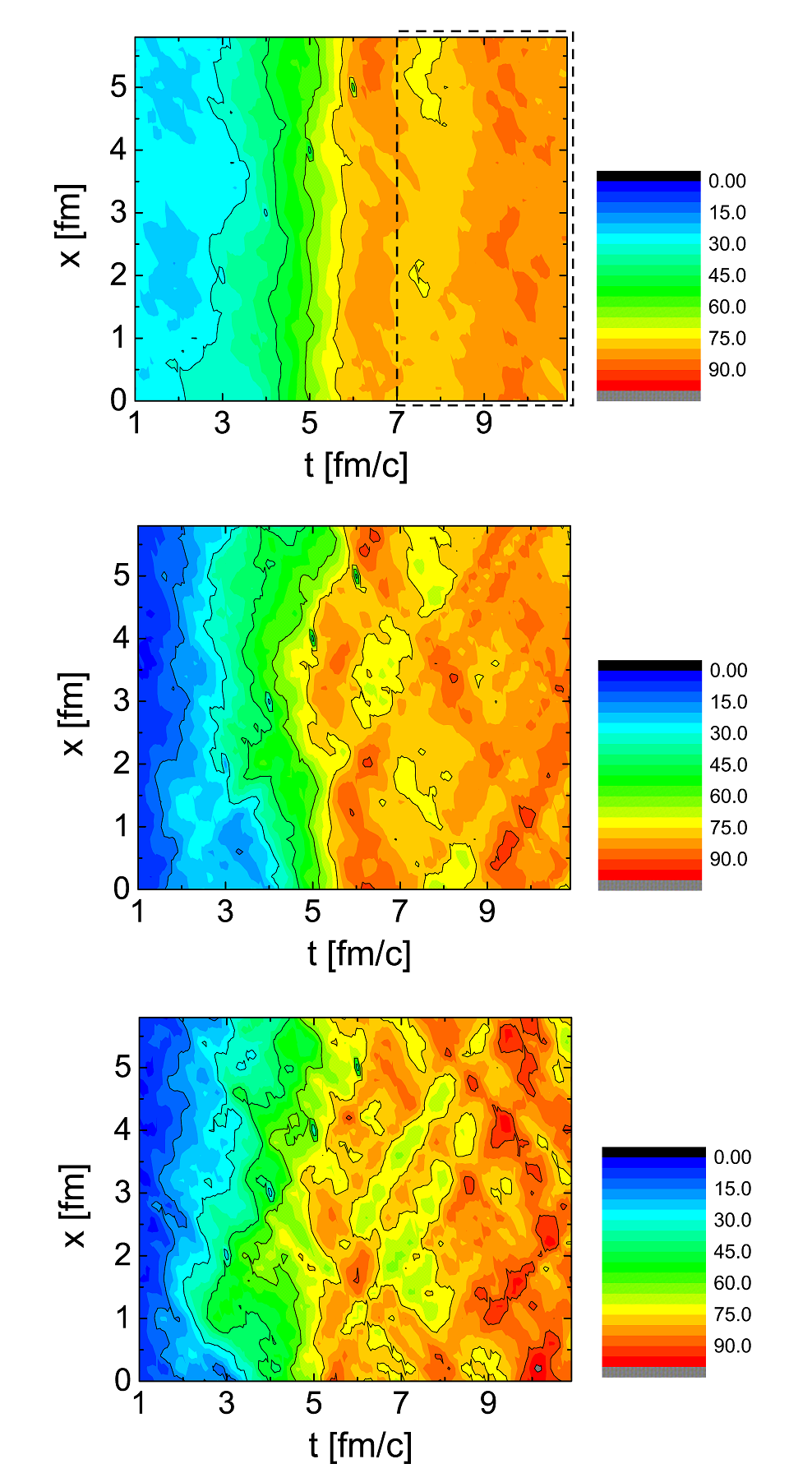}
%\raisebox{0.55\height}{\includegraphics[width=0.4 in]{colorbar.eps}}
%\includegraphics[width=2.5 in]{b=2-3d.eps}
%\raisebox{0.55\height}{\includegraphics[width=0.4 in]{colorbar.eps}}
%\includegraphics[width=2.5 in]{b=3-3d.eps}
%\raisebox{0.55\height}{\includegraphics[width=0.4 in]{colorbar.eps}}
\caption{Dynamical evolution of the $\sigma$ field in space-time coordinate with noise term $B=1, 2, 3$, separately (from up to down). The color bar listed at right hand side represents the strength of the field. The inputs to the CNNs are marked by the dashed box shown in the top figure which applies to every configurations.}
\label{fig:noise}
\end{figure}
%%%%%%%%%%%%%%%%%%%%%%%%%%%%%%%%%%%%%%%%%%%%%%%%%%%%%%%%%%%%%%%%%%%%%%%%%%%%%%

%To perform supervised training, the binary categorical cross-entropy is minimized with the Adam optimizer.
% The initial learning rate is set to be 0.00001.
% might show a CNN structure we used

%To study the higher-level temporal representations of the Langevin dynamics, we use two Long Short-Term Memory (LSTM) layers, which the activation function is hyperbolic tangent. The activation function for the recurrent step is hard sigmoid. (Dropout??)
%For the realistic training and learning of the neural network, we minimize the categorical crossentropy by using the "adam" optimizer. The initial learning rate is set to be 0.00001.
%The structure of our deep neural network is presented in Fig. \ref{rnnstructure}.
%\begin{figure*}[tbp]
%\center
%\includegraphics[width=4.7 in]{RNNStructure}
%\caption{The structure of our Recurrent neural network.}
%\label{rnnstructure}
%\end{figure*}
To perform supervised learning for the binary classification task here, configurations labeled with typical QCD phase transition types (i.e. crossover and first-order) serve as the nuts-and-bolts.
To prepare the two categories of configurations, two trajectories in the $T$-$\mu$ phase diagram are adopted for the dynamical cooling process: the temperature decreases in a Hubble-like way shown in Eq.~\eqref{eq:temp} with fixed baryon chemical potential as $\mu = 180$ MeV and $\mu = 240$ MeV, separately. Both of the two trajectories would experience the phase transition, which provides a homogeneous description to the dynamical evolution. Since the initial configurations of the $\sigma$ field are different in the two trajectories, the one with $\mu = 180$ MeV mimics the crossover transition type and is labelled as $(0, 1)$ in the training set, while the other one at $\mu = 240$ MeV mimics the first-order phase transition and is labelled as $(1, 0)$. The $\sigma$ field evolution processes are recorded as images with size $40\times 30$, where $N_t = 40$ is time grids and $N_s = 30$ is spatial grids. It's worth to note that, the white noise introduced in Eq.~\eqref{eq:langevin} leads to intrinsic differences in the event-by-event generated configurations. The corresponding damping coefficient is set to be $\eta = 1$ fm$^{-1}$ in this section.

For preparing data sets, we simulated $10,000$ events at each parameter setup we considered in our following tasks. Typical $\sigma$ field time evolution are demonstrated in Fig.~\ref{fig:potential}, with two different phase transition scenarios showed. Note that the phase order could not be naively recognized by eye from the spatial-averaged time evolution shown in the figure, much less provided only the final episode evolution after phase transition happening. As a matter of fact, the effective inputs we adopt for the deep CNNs are demonstrated in Fig.~\ref{fig:noise}, i.e. typical $\sigma$ field configurations with final episode masked to be the input. They are 2-dimensional images containing the spatio-temporal information for the evolution, which thus embed more correlations from both the time and space of the dynamics. Different magnitude of noise $B$ are presented in the typical configurations shown in Fig.~\ref{fig:noise}, where $B$ is a free parameter being encoded in the noise term to control the white noise strength. This also eliminate the unreasonably reliance of the noise on the spatial grid size. As enlarging the magnitude of the noise from $B = 1$ to $B = 3$, the configurations become jouncy since the fluctuations is increasing for the dynamical evolution.

%%%%%%%%%%%%%%%%%%%%%%%%%%%%%%%%%%%%%%%%%%%%%%%%%%%%%%%%%%%%%%%%%%%%%%%%%%%%%%
\begin{figure}[htbp!]
\center
\includegraphics[width=3.0 in]{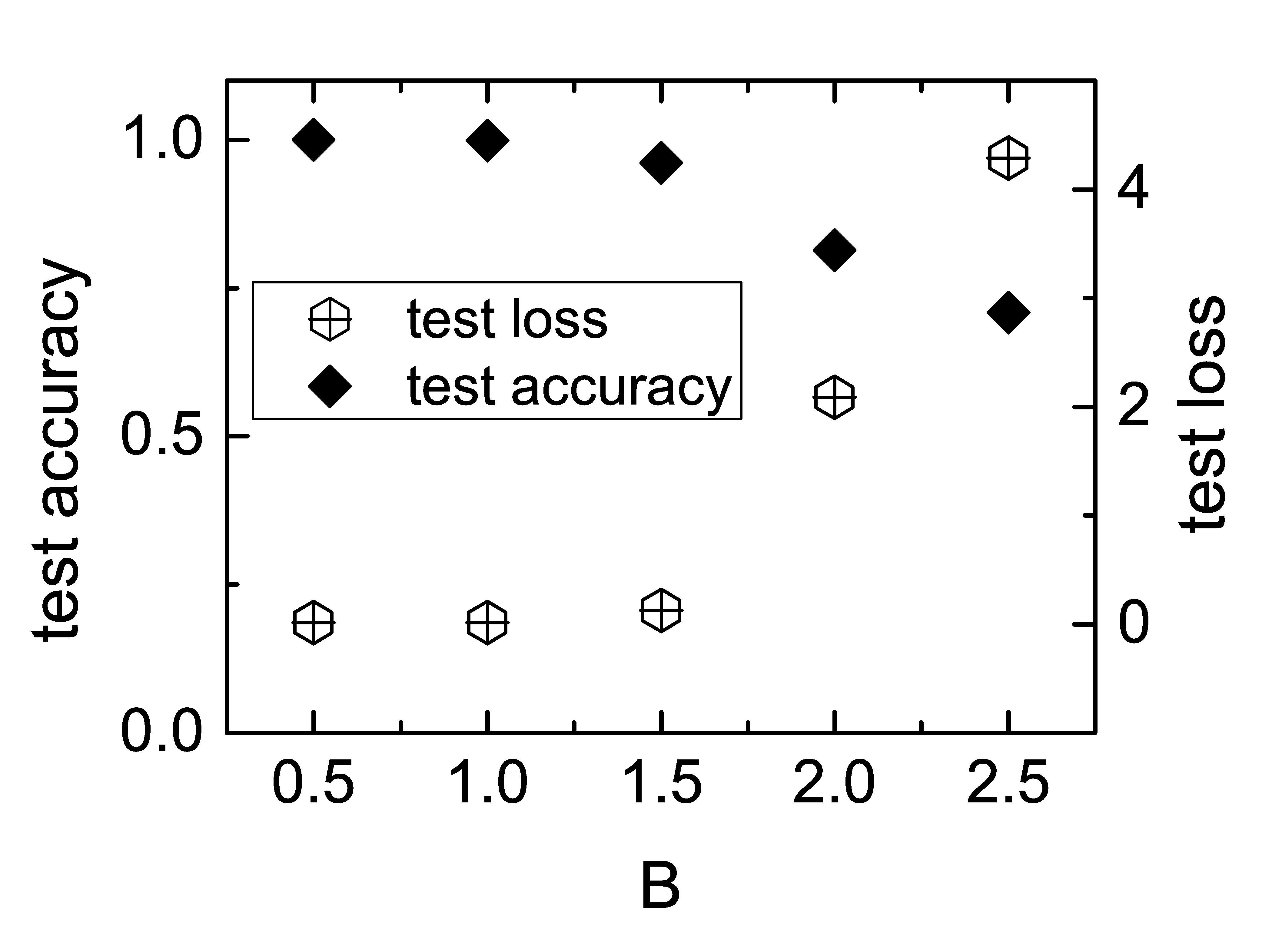}
\caption{The accuracy and loss on test data sets with different noise parameters $B$=0.5, 1, 1.5, 2 and 2.5. The performance is snapped at training epoch=1000.}
\label{fig:acc}
\end{figure}
%%%%%%%%%%%%%%%%%%%%%%%%%%%%%%%%%%%%%%%%%%%%%%%%%%%%%%%%%%%%%%%%%%%%%%%%%%%%%%

In following for applying CNN to the phase transition identification task, from the generated events, $20\%$ of the events are randomly chosen as the test set, and the left part of the data is used for training the neural network. Thus the training data set consists of $20,000$ events with configurations at $B=0.5$ and $20,000$ at $B=1$ (half with first-order half with crossover phase transition), which are fed to the neural network with batch size $=16$. The training runs 1000 epochs, after which the validation accuracy reaches to $99.9\%$. With the trained CNN we further make the test on previously unseen data set. As shown in Fig.\ref{fig:acc}, the deep CNN has an extraordinary performance on recognizing the order of the phase transition. It is worth noting that the input to the network is solely the final episode of the $\sigma$ field configuration after phase transition. The test accuracy is 99\% for configurations with $B=0.5$ and $B=1$, implying the existence of phase order information projected onto the final episode configurations from the dynamical evolution. We also trained a different neural network with LSTM layers to the task based on the same data sets, but the test accuracy reached only 68.7\%. Furthermore, as shown in Fig.~\ref{fig:acc}, the trained CNN could recognize  with above $70\%$ accuracy the phase order information from the configurations which has quite different noise magnitude as included in the training data set. Notably, for $B=1.5$, although the fluctuations are already so large
that can eventually break the deterministic evolution, the test accuracy achieves $95\%$.
It means that the neural network have learned the underlying evolutionary patterns of the phase transition. Even though the $\sigma$ configurations are totally different from the learned ones due to the intense noise, the machine could keep an
accurate prediction on the phase order in each event. In the case of larger noise $B=2$ or $B=2.5$, the machine's performance decreased, which could be ameliorated by introducing more diverse configurations inside the train set to the deep CNN.

\section{Tracking dynamical parameters from evolution}
\label{sec:dynamics}
%%%%%%%%%%%%%%%%%%%%%%%%%%%%%%%%%%%%%%%%%%%%%%%%%%%%%%%%%%%%%%%%%%%%%%%%%%%%%%
\begin{figure}[tbp]
\centering
\includegraphics[width=0.99\linewidth]{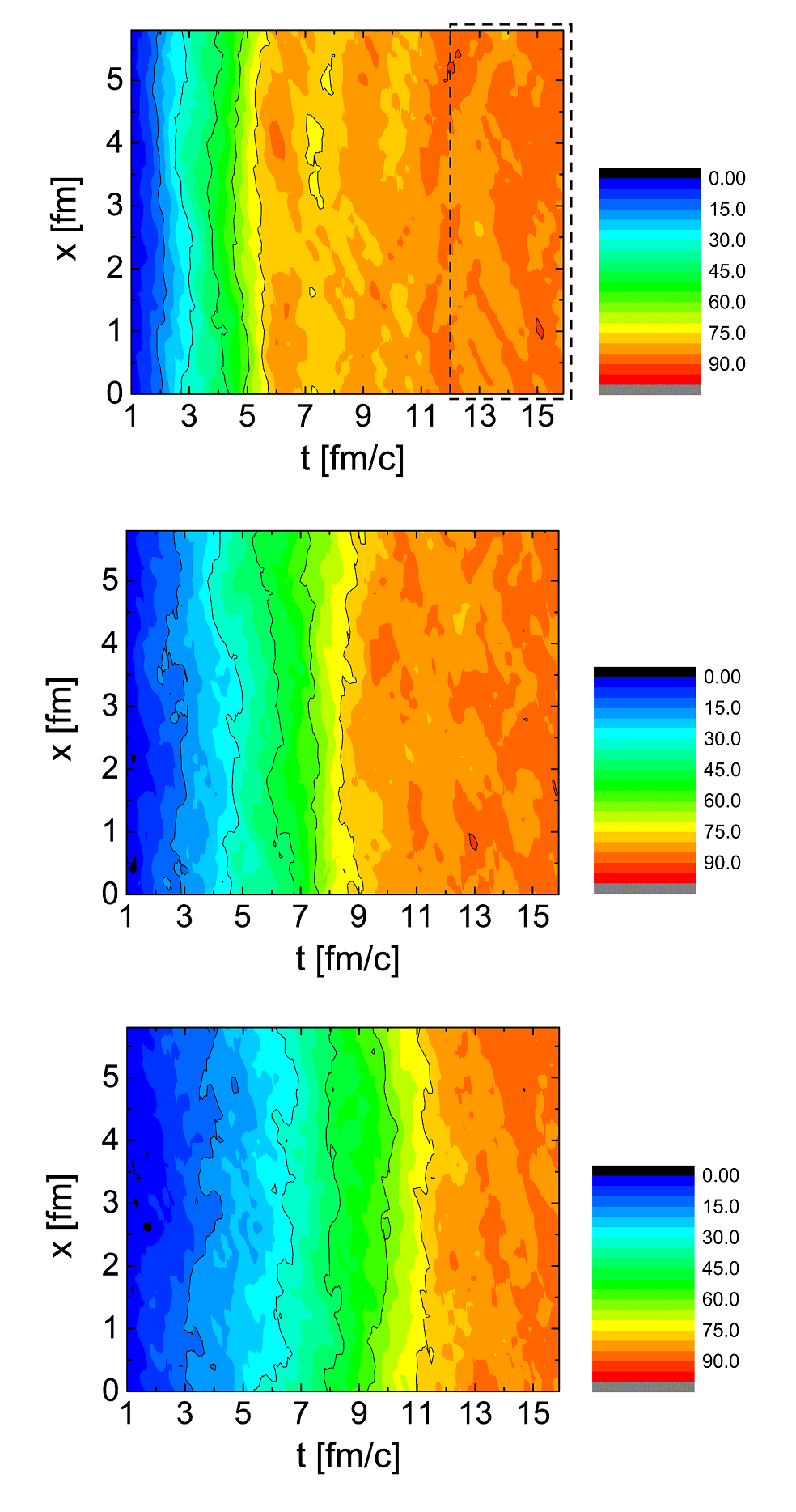}
\caption{Dynamical evolution of the $\sigma$ field in space-time coordinate with the damping coefficient $\eta=1, 3, 5$ fm$^{-1}$, separately (from up to down). The color bar listed at right hand side represents the strength of the field. The inputs to the CNNs are marked by the dashed box shown in the top figure which applies to every configurations.}
\label{fig:eta}
\end{figure}
%%%%%%%%%%%%%%%%%%%%%%%%%%%%%%%%%%%%%%%%%%%%%%%%%%%%%%%%%%%%%%%%%%%%%%%%%%%%%%
In Eq.~\eqref{eq:langevin}, the damping coefficient $\eta$ drives the diffusion process for the $\sigma$ field as a main dynamical parameter. In general, the $\sigma$ field eventually reaches to its vacuum expectation value faster in a smaller damping environment. However, the damping coefficient can influence not only the thermal diffusion speed of the system, but also the fluctuations due to the Einstein's relation. Because of its ambiguity, it is hard to measure the $\eta$ in quantum dynamical systems, no matter whether it's the cold atoms or hot dense Quark-Gluon-Plasma(QGP). In this section, with a well-trained deep CNN,
the good testing results indicate that the damping coefficient could be extracted from the bumpy field configurations collected as the "final episode states" of the stochastic process. To deploy supervised learning on the damping coefficient regression task, we employ similar training strategy as used above for the phase transition binary classification. Specifically, for preparing the training data sets, we used events generated under the Langevin equation on the crossover side with the damping coefficient $\eta$ in the range of  $(1.0-2.5)~\text{fm}^{-1}$ and $(4.6-5.5)~\text{fm}^{-1}$, where $\eta$ varies with size $d\eta = 0.1$. Within each $\eta$ bin, $1,000$ events are generated. Three representative examples of the evolution with different $\eta$ values are shown in Fig.~\ref{fig:eta}, in which there are no distinctive features to recognize different $\eta$. To reduce the bias in data sets, we mix and shuffle the configurations with different $\eta$ values, and randomly divide them into two parts: $20\%$ for validation, the left $80\%$ configurations for training. To investigate the generalization ability of the machine, configurations with different damping coefficient values beyond the training set will be utilized to test the network prediction performance.

%%%%%%%%%%%%%%%%%%%%%%%%%%%%%%%%%%%%%%%%%%%%%%%%%%%%%%%%%%%%%%%%%%%%%%%%%%%%%%
\begin{figure}[htbp!]
\center
\includegraphics[width=3.0 in]{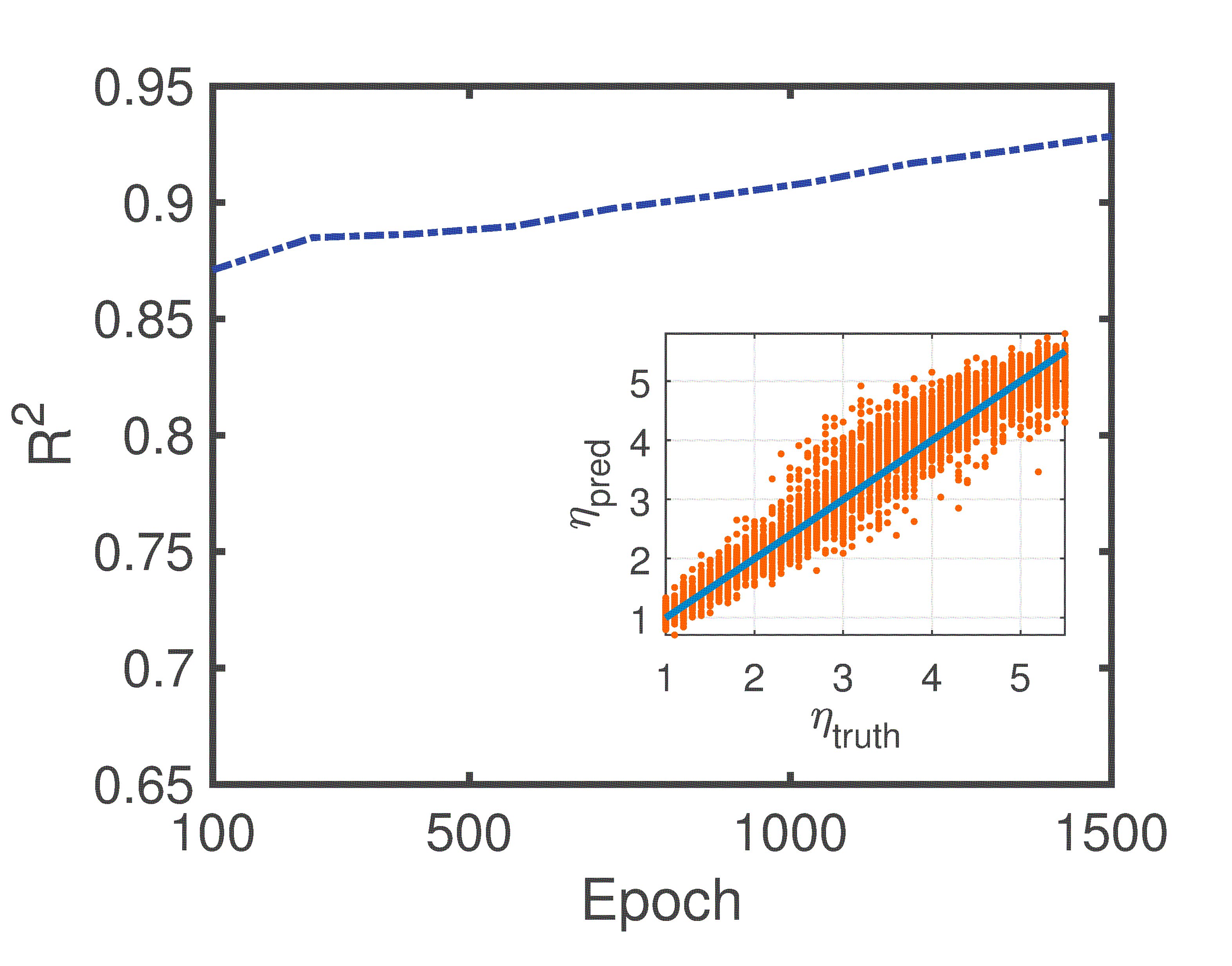}
\caption{The training process in predicting the damping coefficient from configurations. The blue dash line in main figure is the training curve which contains the coefficient of determination $R^2$ increasing with epochs. The insert part consists of 4600 orange dots which are labeled by the damping coefficients of the ground truth and predictions from the trained CNNs.}
\label{fig:test}
\end{figure}
%%%%%%%%%%%%%%%%%%%%%%%%%%%%%%%%%%%%%%%%%%%%%%%%%%%%%%%%%%%%%%%%%%%%%%%%%%%%%

%The validation accuracy reaches $99.3 \%$.
A network with similar architecture as shown in Fig.~\ref{fig:flowchart} is deployed, to which slight changes are performed to target this regression task. Specifically, the Dropout layers are removed, and one more fully connected layer with 32 neurons is inserted before the final output layer. The latter is with one neuron representing the values of $\eta$. The activation function are all changed to Parameterized ReLU and the loss function is set to be the mean squared error between the true values and network predictions. With that, the training process is depicted in Fig.~\ref{fig:test}, where the coefficient of determination $R^2$ are calculated to indicate the regression quality,
\begin{equation}
R^2=1 - \frac{SS_\text{res}}{SS_\text{tot}},
\end{equation}
with $SS_\text{res}=\sum_i(\eta_{i,\text{truth}}-\bar{\eta}_\text{truth}), SS_\text{tot}=\sum_i(\eta_{i,\text{truth}}-\eta_{i,\text{pred}})$ are the residual sum of squares and the total sum of squares, respectively. The $R^2$ is implemented to quantitatively evaluate the correlation between the ground truth and the prediction from the neural networks. As training goes on, the $R^2$ grows quickly from $87\%$ to $93\%$ and tends to be stable, which reveals that the deep CNNs has advantages to capture the hidden correlation in image-type inputs~\cite{levine:2019quantum}. With regard to the insert figure in Fig.~\ref{fig:test}, the trained regression CNN is tested on previously unseen $\sigma$ field configurations with different values of the damping coefficients. The results demonstrate that the predictions retain high-consistency with the value of the damping coefficient for diverse configurations with stochasticity. The predicted $\eta$ values versus the corresponding ground-truth lies around the diagonal line with a band indicating the deviation. Remarkably, it is found that although inside the training set there's no supervision in the region of $\eta\sim(2.6-4.5)~\text{fm}^{-1}$, the predictions of the trained network still keeps a reasonable performance in this range. In such interpolation region, the large damping will induce fluctuations definitely non-negligible for the evolution, which makes it hard to decode the dynamical parameters via any conventional analysis. The deep learning approach in our work offers an alternative way to track the dynamics of stochastic process from intricate configurations driven by the damping coefficient.

\section{summary and outlook}
\label{sec:discus}
In this paper, we introduced a method applying deep learning to identify the phase transition information and also track the dynamics for a stochastic dynamical models near QCD critical point. We numerically simulate the time evolution of the fluctuating $1+1$-dimensional $\sigma$ field within the framework of Langevin dynamics on event-by-event manner, and collect its spatial-temporal field configurations to form the data sets for the deep learning study on our tasks.

Based on the generated data, the machine can be trained to identify the nature of the phase transition: first-order or crossover type, encoded inside the stochastic field dynamical evolution. Although the field configurations are totally different from each other due to the noise terms and also the random initial conditions, the machine successfully learns to make accurate prediction on the phase order for previously unseen evolution events in testing stage. This is related to the powerful capability of the deep CNNs for extracting hidden correlations in image-type data-set, which facilitates the presented phase order identification from the field configurations. We further designed a regressive CNN to decode the dynamical parameter -- damping coefficient -- inside the Langevin dynamics from the evolution.
%Additionally, we prepare mixed configurations with damping coefficient in the range of $(1.0-2.5)~fm^{-1}$ and $(4.6-5.5)~fm^{-1}$, nevertheless, the predictions to the dynamics are made for the entire range.
It is found that extracting dynamics from such stochastic process with the trained CNN shows robustness, which also reveals an acceptable generalization ability when tested on configurations containing damping beyond the training set. In summary, we demonstrate that the framework is effective in extracting the Langevin dynamics from complicated configurations associated with intrinsic stochasticity.

The present method can be helpful for a broader field, like, there is a potential application in topological-dependent stochastic process~\cite{zhao:2020topologydependent}, in which the topological charge could be extracted by the deep CNNs in a similar manner. Moreover, the transfer learning could also help us to understand the stochastic process through introducing the well-trained deep CNNs
into real physical observations~\cite{kwon:2020magnetic}.

\section*{Acknowledgment}
The authors acknowledge inspiring discussions with Horst Stoecker.
The work is supported by the AI grant at FIAS of SAMSON AG, Frankfurt (L. J., L. W. and K. Z.), by the BMBF under the ErUM-Data project (K. Z.),  and by the NVIDIA Corporation with the generous donation of NVIDIA GPU cards for the research (K. Z.).

%\bibliographystyle{unsrt}
%\section*{Bibliography}
%\bibliographystyle{apsrev4-1}
% \bibliography{jiang}
%\bibliography{langevin}

\end{document}